\documentclass[prl,unsortedaddress,superscriptaddress,showpacs,twocolumn,floatfix]{revtex4}

\usepackage{graphicx}% Include figure files

\begin{document}

\title{Measurement of the Michel Parameter $\rho$ in Muon Decay}

\affiliation{University of Alberta, Edmonton, AB, T6G 2J1, Canada}
\affiliation{University of British Columbia, Vancouver, BC, V6T 1Z1, Canada}
\affiliation{Kurchatov Institute, Moscow, 123182, Russia}
\affiliation{University of Montreal, Montreal, QC, H3C 3J7, Canada}
\affiliation{University of Regina, Regina, SK, S4S 0A2, Canada}
\affiliation{Texas A\&M University, College Station, TX 77843, U.S.A.}
\affiliation{TRIUMF, Vancouver, BC, V6T 2A3, Canada}
\affiliation{Valparaiso University, Valparaiso, IN 46383, U.S.A.}

\author{J.R.~Musser}
\affiliation{Texas A\&M University, College Station, TX 77843, U.S.A.}

\author{R.~Bayes}
\altaffiliation[Affiliated with: ]{Univ.\@ of Victoria,
Victoria, BC.}
\affiliation{TRIUMF, Vancouver, BC, V6T 2A3, Canada}

\author{Yu.I.~Davydov}
\altaffiliation[Affiliated with: ]{Kurchatov Institute,
Moscow, Russia.}
\affiliation{TRIUMF, Vancouver, BC, V6T 2A3, Canada}

\author{P.~Depommier}
\affiliation{University of Montreal, Montreal, QC, H3C 3J7, Canada}

\author{J.~Doornbos}
\affiliation{TRIUMF, Vancouver, BC, V6T 2A3, Canada}

\author{W.~Faszer}
\affiliation{TRIUMF, Vancouver, BC, V6T 2A3, Canada}

\author{C.A.~Gagliardi}
\affiliation{Texas A\&M University, College Station, TX 77843, U.S.A.}

\author{A.~Gaponenko}
\affiliation{University of Alberta, Edmonton, AB, T6G 2J1, Canada}

\author{D.R.~Gill}
\affiliation{TRIUMF, Vancouver, BC, V6T 2A3, Canada}

\author{P.~Green}
\affiliation{University of Alberta, Edmonton, AB, T6G 2J1, Canada}

\author{P.~Gumplinger}
\affiliation{TRIUMF, Vancouver, BC, V6T 2A3, Canada}

\author{M.D.~Hasinoff}
\affiliation{University of British Columbia, Vancouver, BC, V6T 1Z1, Canada}

\author{R.S.~Henderson}
\affiliation{TRIUMF, Vancouver, BC, V6T 2A3, Canada}

\author{J.~Hu}
\affiliation{TRIUMF, Vancouver, BC, V6T 2A3, Canada}

\author{B.~Jamieson}
\affiliation{University of British Columbia, Vancouver, BC, V6T 1Z1, Canada}

\author{P.~Kitching}
\affiliation{University of Alberta, Edmonton, AB, T6G 2J1, Canada}

\author{D.D.~Koetke}
\affiliation{Valparaiso University, Valparaiso, IN 46383, U.S.A.}

\author{A.A.~Krushinsky}
\affiliation{Kurchatov Institute, Moscow, 123182, Russia}

\author{Yu.Yu.~Lachin}
\affiliation{Kurchatov Institute, Moscow, 123182, Russia}

\author{J.A.~Macdonald}
\altaffiliation[Deceased.]{} 
\affiliation{TRIUMF, Vancouver, BC, V6T 2A3, Canada}

\author{R.P.~MacDonald}
\affiliation{University of Alberta, Edmonton, AB, T6G 2J1, Canada}

\author{G.M.~Marshall}
\affiliation{TRIUMF, Vancouver, BC, V6T 2A3, Canada}

\author{E.L.~Mathie}
\affiliation{University of Regina, Regina, SK, S4S 0A2, Canada}

\author{L.V.~Miasoedov}
\affiliation{Kurchatov Institute, Moscow, 123182, Russia}

\author{R.E.~Mischke}
\affiliation{TRIUMF, Vancouver, BC, V6T 2A3, Canada}

\author{P.M.~Nord}
\affiliation{Valparaiso University, Valparaiso, IN 46383, U.S.A.}

\author{K.~Olchanski}
\affiliation{TRIUMF, Vancouver, BC, V6T 2A3, Canada}

\author{A.~Olin}
\altaffiliation[Affiliated with: ]{Univ.\@ of Victoria,
Victoria, BC.}
\affiliation{TRIUMF, Vancouver, BC, V6T 2A3, Canada}

\author{R.~Openshaw}
\affiliation{TRIUMF, Vancouver, BC, V6T 2A3, Canada}

\author{T.A.~Porcelli}
\altaffiliation[Present address: ]{Univ.\@ of Manitoba,
Winnipeg, MB.}
\affiliation{TRIUMF, Vancouver, BC, V6T 2A3, Canada}

\author{J.-M.~Poutissou}
\affiliation{TRIUMF, Vancouver, BC, V6T 2A3, Canada}

\author{R.~Poutissou}
\affiliation{TRIUMF, Vancouver, BC, V6T 2A3, Canada}

\author{M.A.~Quraan}
\affiliation{University of Alberta, Edmonton, AB, T6G 2J1, Canada}

\author{N.L.~Rodning}
\altaffiliation[Deceased.]{} 
\affiliation{University of Alberta, Edmonton, AB, T6G 2J1, Canada}

\author{V.~Selivanov}
\affiliation{Kurchatov Institute, Moscow, 123182, Russia}

\author{G.~Sheffer}
\affiliation{TRIUMF, Vancouver, BC, V6T 2A3, Canada}

\author{B.~Shin}
\altaffiliation[Affiliated with: ]{Univ.\@ of Saskatchewan,
Saskatoon, SK.}
\affiliation{TRIUMF, Vancouver, BC, V6T 2A3, Canada}

\author{F.~Sobratee}
\affiliation{University of Alberta, Edmonton, AB, T6G 2J1, Canada}

\author{T.D.S.~Stanislaus}
\affiliation{Valparaiso University, Valparaiso, IN 46383, U.S.A.}

\author{R.~Tacik}
\affiliation{University of Regina, Regina, SK, S4S 0A2, Canada}

\author{V.D.~Torokhov}
\affiliation{Kurchatov Institute, Moscow, 123182, Russia}

\author{R.E.~Tribble}
\affiliation{Texas A\&M University, College Station, TX 77843, U.S.A.}

\author{M.A.~Vasiliev}
\affiliation{Texas A\&M University, College Station, TX 77843, U.S.A.}

\author{D.H.~Wright}
\altaffiliation[Present address: ]{Stanford Linear Accelerator Center,
Stanford, CA.}
\affiliation{TRIUMF, Vancouver, BC, V6T 2A3, Canada}

\collaboration{TWIST Collaboration}
\noaffiliation

\date{\today}

\begin{abstract}
The TWIST Collaboration has measured the Michel parameter $\rho$ in normal muon decay, $\mu^+ \to e^+ \nu_e \bar{\nu}_{\mu}$.  In the Standard Model, $\rho$ = 3/4.  Deviations from this value require mixing of left- and right-handed muon and electron couplings in the muon-decay Lagrangian.  We find $\rho$ = 0.75080 $\pm$ 0.00044(stat.) $\pm$ 0.00093(syst.) $\pm$ 0.00023, where the last uncertainty represents the dependence of $\rho$ on the Michel parameter $\eta$.  This result sets new limits on the $W_L-W_R$ mixing angle in left-right symmetric models.
\end{abstract}

\pacs{13.35.Bv, 14.60.Ef, 12.60.Cn}% PACS codes
                                    % Use showpacs class option to display
\maketitle

Normal muon decay, $\mu^+ \to e^+ \nu_e \bar{\nu}_{\mu}$, is an excellent laboratory to test the space-time structure of the weak interaction.  The energy and angular distributions of the positrons emitted in the decay of polarized muons can be described in terms of four parameters -- $\rho$, $\eta$, $\xi$, and $\delta$ -- commonly referred to as the Michel parameters.  Neglecting the electron and neutrino masses and radiative corrections, the differential decay rate for positive muon decay is given in terms of $\rho$, $\xi$, and $\delta$ by \cite{Michel_decay}:
\begin{eqnarray}
\frac{d^2 \Gamma}{x^2 dx d(\cos \theta)} & \propto & (3 - 3x) + \frac{2}{3} \rho (4x-3)
\nonumber \\
& + & P_{\mu} \xi \cos\theta \left[ (1-x) + \frac{2}{3} \delta (4x-3) \right],
\label{eq:michel}
\end{eqnarray}
where $P_{\mu}$ is the polarization of the muon, $x$ is the outgoing positron energy as a fraction of the maximum possible value, and $\theta$ is the angle between the muon polarization axis and the positron decay direction.  The fourth decay parameter, $\eta$, contributes to the angle-independent part of the distribution if one includes the finite electron mass.  In this Letter, the TWIST Collaboration reports a new measurement of the Michel parameter $\rho$.  A concurrent measurement of the parameter $\delta$ is described in Ref.\@ \cite{Delta04}.

The current accepted value of $\rho$, 0.7518 $\pm$ 0.0026 \cite{Peop66,PDG}, is consistent with the Standard Model expectation, $\rho$ = 3/4.  Any deviation from 3/4 would imply the muon-decay Lagrangian includes scalar, vector, or tensor couplings between left-handed muons and right-handed electrons or vice versa \cite{Fets86}.  For example, in left-right symmetric models,
the $W_L-W_R$ mixing angle $\zeta$ is given by \cite{Herz86}
\begin{equation}
\zeta = \sqrt{ \frac{2}{3} \left(\frac{3}{4} - \rho\right) }.
\end{equation}
Unlike many other limits on right-handed currents, this is independent of the form of the right-handed CKM matrix.  Recently, $\rho$ has also been related to loop corrections to the neutrino mass matrix \cite{Prez04}.  For a review of muon decay within the Standard Model, see Ref.\@ \cite{Kuno01}.

TWIST utilizes the M13 beam line at TRIUMF to transport beams of 29.6 MeV/c surface muons from pion
decay-at-rest ($P_{\mu}$\,$\sim$\,$-1$) or 32.8 MeV/c cloud muons from pion
decay-in-flight ($P_{\mu}$\,$\sim$\,$+0.25$) into the TWIST spectrometer.  The TWIST spectrometer consists of an array of very thin, high precision planar wire chambers located within a 2-T magnetic field oriented along the beam direction.  The spectrometer includes 44 drift chamber (DC) planes operated with DME gas and 12 multi-wire proportional chamber (PC) planes operated with CF$_4$-isobutane (80-20) gas, all within a He-N$_2$ (97-3) enclosure.  The wire planes are symmetrically located upstream and downstream of a 125-$\mu$m thick Mylar stopping target, with $10^{+10}_{-5}$ $\mu$m of graphite painted on each surface.
A detailed description of the TWIST spectrometer is given in Ref.\@ \cite{NIM04}.

After muons enter the magnetic field, they pass through a thin plastic scintillator that provides the event trigger.  They then pass through the detector planes until they stop in the target.  Decay positrons follow helical trajectories through the DCs and PCs, permitting their momenta and decay angles to be measured precisely.  For each event, all DC and PC hits within an interval from 6 $\mu$s before until 10 $\mu$s after the trigger time are recorded.

During off-line analysis, the PC and DC hits are examined to identify events in which the muon stopped in the target, then decayed at least 1.05 $\mu$s, and no more than 9 $\mu$s, later.  The delay insures that the PC and DC hits associated with the muon and decay positron do not overlap.  Events are rejected if a second muon enters the spectrometer, or if a beam positron passes through the spectrometer within 1.05 $\mu$s of either the muon arrival or decay time.  Additional cuts include the muon flight time through the M13 beam line and a requirement that the muon stopping location be within 2.5 cm of the detector axis.  All events that pass these cuts are analyzed to reconstruct the decay positron kinematics.

After track fitting, 2.3\% of the events contain additional tracks in coincidence with the decay.  Extra tracks can arise from beam particles that are not resolved in time, events that scatter within the detector leading to two reconstructed track segments, and events that include delta rays or decay positrons that backscatter from material outside the detector volume.  Two algorithms have been developed to select among the choices in multi-track events.  They also impose different constraints on events that scatter within the detector when only one track segment is reconstructed.  All events have been analyzed using both algorithms.

To extract the Michel parameters, the measured positron spectrum is compared to that predicted by a detailed Monte Carlo (MC) simulation.  The MC uses GEANT \cite{geant} to simulate particle interactions and a model based on GARFIELD \cite{garfield} to simulate wire chamber responses.  The MC decay generator includes the effects of electron mass, plus first-order and many higher-order radiative corrections \cite{Rad_corr}.  It also includes beam positrons and additional muons in the simulated events according to their observed rates in the data.  The output from the MC is digitized and processed by the same analysis codes that are used for real events.

The data reported here include a total of $6 \times 10^9$ muon decay events that were recorded during Fall, 2002.  Sixteen independent data sets were taken to explore the sensitivity of the spectrometer and analysis to a broad range of systematic effects.
A typical data set included $3 \times 10^8$ events, sufficient to determine $\rho$ with a statistical precision of $\sim$\,0.0007.  In addition, special runs were taken to provide data to validate aspects of the simulation that are difficult to test with the muon decay spectrum.
Five data sets were taken under conditions that permit a reliable determination of $\rho$.  Four sets were taken with a surface muon beam.  Sets A and B were obtained six weeks apart at a magnetic field of 2.00 T; the other sets were taken at 1.96 T and 2.04 T.  The fifth data set was taken at 2.00 T with a cloud muon beam.
Depolarizing interactions in the target reduced the average muon polarization at the time of decay to $\sim$\,90\% of the incident polarization.

\begin{figure}[tb]
\center{\includegraphics*[width=8cm]{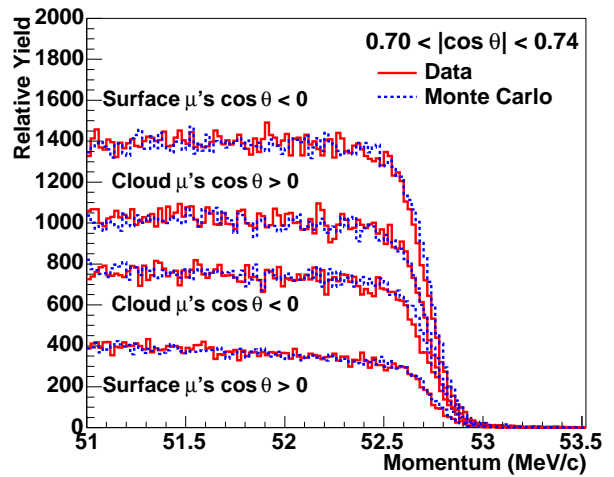}}
\caption{(color online)  Measured positron momentum spectra (solid lines) near the end point are compared to Monte Carlo simulations (dotted lines).  The curves show surface muon set B for $-0.74 < \cos\theta < -0.70$, cloud muons for $+0.70 < \cos\theta < +0.74$, cloud muons for $-0.74 < \cos\theta < -0.70$, and surface muon set B for $+0.70 < \cos\theta < +0.74$.}
\label{fig:endpoint}
\end{figure}

The only discrete feature in the muon decay spectrum is the end point.  Figure \ref{fig:endpoint} shows comparisons of the measured spectra near the end point to MC simulations.
The observed end point falls below the kinematic limit of 52.828 MeV/c due to positron energy loss in the target and detector materials.  Fits to spectra in the region $p > 52$ MeV/c for various decay angles show that the energy loss follows the form, $\Delta E(\theta) =
-\alpha/|\cos\theta|$, with $\alpha$ a constant as expected for the planar geometry of the TWIST detectors \cite{NIM97}.  $\alpha$ takes on different values for upstream and downstream decays when the muon stopping distribution is not centered in the target.
We use $E = E_{meas} + \alpha/|\cos\theta|$ to correct both data and MC events for the average positron energy loss.

\begin{figure}[tb]
\center{\includegraphics*[width=7.5cm]{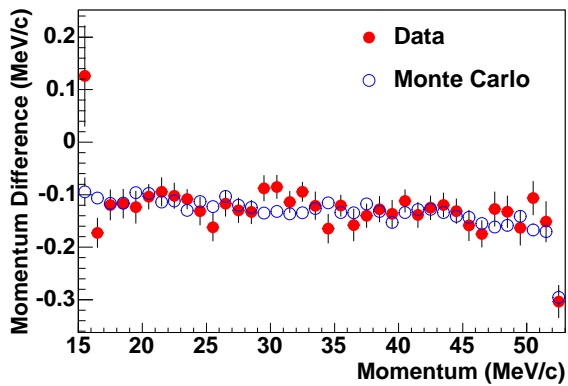}}
\caption{(color online)  Average energy loss of decay positrons through the target and detector materials as a function of momentum for data (closed circles) and Monte Carlo (open circles), measured as described in the text.}
\label{fig:eloss}
\end{figure}

Effects that distort the reconstructed positron momenta will lead to systematic errors in the extracted value of $\rho$ if they are not simulated accurately by the MC.  To test the simulation of energy loss for positron momenta well below 52.83 MeV/c, events were recorded in which a muon came to rest at the far upstream end of the detector.  Positrons from muon decays in the downstream direction first spiral through the upstream half of the detector, then pass through the target, and finally spiral through the downstream half of the detector.  Figure \ref{fig:eloss} shows the difference between the reconstructed positron momenta in the two halves, which measures the energy loss in the target and detector materials.  The MC agrees very well with the data.  Similar comparisons verify the MC simulation of positron multiple scattering and hard interaction rates.

If $\rho = \rho_H + \Delta\rho$ and $\eta = \eta_H + \Delta\eta$, then the angle-integrated muon decay spectrum can be written as:
\begin{equation}
N(x) = N_S(x,\rho_H,\eta_H) + \Delta\rho N_{\Delta\rho}(x) + \Delta\eta N_{\Delta\eta}(x).
\end{equation}
The spectrum is linear in the decay parameters, so this expansion is exact.  It can also be generalized to include the angular dependence.  This is the basis for the blind analysis.
The measured momentum-angle spectrum is fitted to the sum of a MC `standard' spectrum $N_S$ produced with unknown Michel parameters $\rho_H$, $\eta_H$, $\xi_H$, $\delta_H$ and additional `derivative' MC distributions $N_{\Delta\rho}$, $N_{\Delta\xi}$, and $N_{\Delta\xi\delta}$, with $\Delta\rho$, $\Delta\xi$, and $\Delta\xi\delta$ as the fitting parameters.  The hidden Michel parameters associated with $N_S$ are revealed only after all data analysis has been completed.  Finally, a refit is performed to shift $\eta$ from the hidden value $\eta_H$ to the accepted value.  The fiducial region adopted for this analysis requires $p<50$ MeV/c, $|p_z|>13.7$ MeV/c, $p_T<38.5$ MeV/c, and $0.50<|\cos\theta|<0.84$.

\begin{figure}[tb]
\center{\includegraphics*[width=8.3cm]{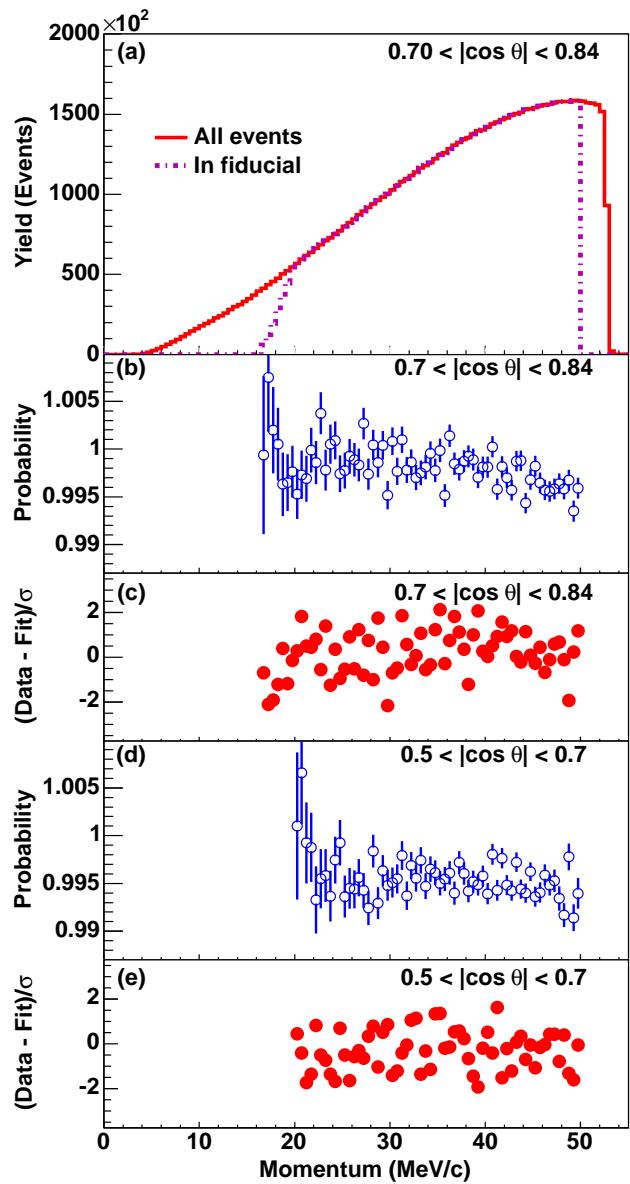}}
\caption{(color online)  Panel (a) shows the muon decay spectrum (solid curve) from surface muon set B as a function of momentum, for events with $0.70 < |\cos\theta | < 0.84$, as well as the events within this angular region that pass the fiducial constraints (dot-dashed curve).  Panels (b) and (d) show the probability for reconstructing muon decays for two angular ranges, as calculated by the Monte Carlo.  Panels (c) and (e) show the residuals for the same angular ranges from the fit of set B to the Monte Carlo `standard' spectrum plus derivatives.}
\label{fig:results}
\end{figure}

Figure \ref{fig:results}(a) shows the momentum spectrum from set B in the angular range $0.70 < |\cos\theta| < 0.84$.  The probability for reconstructing muon decays is very high, as shown in Figs.\@ \ref{fig:results}(b) and (d).
Thus, higher momentum decays that undergo hard interactions and are reconstructed at lower momenta can lead to an apparent reconstruction probability above unity.  Figures \ref{fig:results}(c) and (e) show the residuals of the fit of the decay spectrum from set B for two different angular regions.  Similar fits have been performed to the other data sets, yielding the results shown in Table \ref{table:results}.

\begin{table}[tb]
\caption{\label{table:results}
Results from the fits to the various data sets using one of the track selection algorithms.  Each fit has 1887 degrees of freedom.  Only statistical and set-dependent systematic uncertainties are shown.}
\begin{ruledtabular}
\begin{tabular}{lcc}
Data Set & $\rho$ & $\chi^2$ \\
\hline
Set A    & 0.75134 $\pm$ 0.00083 $\pm$ 0.00053 & 1814 \\
Set B    & 0.74937 $\pm$ 0.00066 $\pm$ 0.00053 & 1965 \\
1.96 T   & 0.75027 $\pm$ 0.00065 $\pm$ 0.00055 & 1951 \\
2.04 T   & 0.75248 $\pm$ 0.00070 $\pm$ 0.00060 & 1804 \\
Cloud    & 0.75157 $\pm$ 0.00076 $\pm$ 0.00053 & 1993 \\
\end{tabular}
\end{ruledtabular}
\end{table}

The 11 additional data sets have been combined with further MC simulations to estimate the systematic uncertainties shown in Tables \ref{table:results} and \ref{table:systs}.  The largest effects arise from time-variations of the cathode foil locations \cite{NIM04} and the density of the DME gas, which change the drift velocities and influence the DC efficiencies far from the sense wires.  These parameters were monitored throughout the data taking, but only average values were used in the analysis.  Special data sets and MC simulations that amplify these effects have been used to estimate their uncertainties for $\rho$.  Other important effects arise from the uncertainty in the thickness of the graphite layers on the Mylar target \cite{NIM04} and from uncertainties in the GEANT treatment of positron interactions that lead to the spectrum distortions seen in Fig.\@ \ref{fig:results}.  Upper limits on these uncertainties were obtained from the data where the muons stopped far upstream.  Smaller effects arise from the uncertainty in the orientation of the detector with respect to the magnetic field, the statistical uncertainties in the end point fits that are used in the momentum calibrations, the theoretical uncertainty in the higher-order radiative corrections, and time-variations in the average muon stopping location within the target.

\begin{table}[tb]
\caption{\label{table:systs}
Contributions to the systematic uncertainty in $\rho$.  Average values
are given for those denoted (ave), which are
considered set dependent when performing the weighted average of the data sets.}
\begin{ruledtabular}
\begin{tabular}{lc}
Effect & Uncertainty \\
\hline
Chamber response (ave)      & $\pm$\,0.00051 \\
Stopping target thickness   & $\pm$\,0.00049 \\
Positron interactions       & $\pm$\,0.00046 \\
Spectrometer alignment      & $\pm$\,0.00022 \\
Momentum calibration (ave)  & $\pm$\,0.00020 \\
Theoretical radiative corrections \protect\cite{Rad_corr} & $\pm$\,0.00020 \\
Track selection algorithm   & $\pm$\,0.00011 \\
Muon beam stability (ave)   & $\pm$\,0.00004 \\
\end{tabular}
\end{ruledtabular}
\end{table}

We treat the chamber response, momentum calibration, and beam stability uncertainties as set dependent when computing the average of the data sets since the underlying causes fluctuated in time.  This gives $\rho = 0.75091 \pm 0.00032$(stat.), with $\chi^2$ = 7.5 for 4 degrees of freedom.  The alternative track selection technique gives $\rho$ = 0.75069.  We average these results as our best estimate of $\rho$, and include half the difference in the systematic uncertainty.  We then rescale the statistical uncertainty by $\sqrt{\chi^2_\nu}$ and, to be conservative, consider the systematic uncertainties as common to the five data sets.

We find $\rho$ = 0.75080 $\pm$ 0.00044(stat.) $\pm$ 0.00093(syst.) $\pm$ 0.00023, consistent with the Standard Model expectation $\rho$ = 3/4.  This result assumes that $\eta$ is given by the accepted value \cite{PDG}, $\eta = -0.007 \pm 0.013$; the third uncertainty represents the change in $\rho$ when $\eta$ changes within its uncertainty.  Within left-right symmetric models, this result sets a new upper limit, $|\zeta|<0.030$ (90\% c.l.), on the $W_L-W_R$ mixing angle.

We acknowledge M.C.\@ Fujiwara and M.\@ Nozar for assistance during the final phase of the data analysis.
We thank P.A.\@ Amaudruz, C.A.\@ Ballard, M.J.\@ Barnes, S.\@ Chan, B.\@ Evans, M.\@ Goyette, K.W.\@ Hoyle, D.\@ Maas, J.\@ Schaapman, J.\@ Soukup, C.\@ Stevens, G.\@ Stinson, H.-C.\@ Walter, and the many undergraduate students who contributed to the construction and operation of TWIST.
We also acknowledge many contributions by other professional and technical staff members from TRIUMF and collaborating institutions.
Computing resources for the analysis were provided by WestGrid.
This work was supported in part by the Natural Sciences and Engineering Research Council and the National Research Council of Canada, the Russian Ministry of Science, and the U.S.\@ Department of Energy.


\begin{thebibliography}{99}

\bibitem{Michel_decay} L. Michel, Proc. Phys. Soc. \textbf{A63}, 514 (1950);
C. Bouchiat and L. Michel, Phys. Rev. \textbf{106}, 170 (1957);
T. Kinoshita and A. Sirlin, Phys. Rev. \textbf{108}, 844 (1957).

\bibitem{Delta04} A. Gaponenko \textit{et al}. (TWIST Collaboration), to be submitted.

\bibitem{Peop66} J. Peoples, Ph.D. thesis, Columbia University, Nevis Report No. 147 (1966), unpublished;
B.A. Sherwood, Phys. Rev. \textbf{156}, 1475 (1967);
D. Fryberger, Phys. Rev. \textbf{166}, 1379 (1968);
S.E. Derenzo, Phys. Rev. \textbf{181}, 1854 (1969).

\bibitem{PDG} S. Eidelman \textit{et al}., Phys. Lett. \textbf{B592}, 1 (2004).

\bibitem{Fets86} W. Fetscher and H.-J. Gerber, \textit{Precision Tests of the Standard Electroweak Model}, ed. by P. Langacker (World Scientific, Singapore, 1995), 657.

\bibitem{Herz86} P. Herczeg, Phys. Rev. D \textbf{34}, 3449 (1986).

\bibitem{Prez04} G. Prezeau, A. Kurylov, and M.J. Ramesy-Musolf, hep-ph/0409193.

\bibitem{Kuno01} Y. Kuno and Y. Okada, Rev. Mod. Phys. \textbf{73}, 151 (2001).

\bibitem{NIM04} R.S. Henderson \textit{et al}., Nucl. Instrum. Methods, submitted; TRIUMF preprint PP-04-20.

\bibitem{geant} GEANT, version 3.2114.

\bibitem{garfield} GARFIELD, version 7.10.

\bibitem{Rad_corr} A.B. Arbuzov, Phys. Lett. \textbf{B524}, 99 (2002); JHEP \textbf{0303}, 063 (2003); JTEP Lett. \textbf{78}, 179 (2003); A. Arbuzov, A. Czarnecki, and A. Gaponenko, Phys. Rev. D \textbf{65}, 113006 (2002); A. Arbuzov and K. Melnikov, Phys. Rev. D \textbf{66}, 093003 (2002).

\bibitem{NIM97} A.A. Khrutchinsky, Yu.Yu Lachin, and V.I. Selivanov, Nucl. Instrum. Meth. \textbf{A396}, 135 (1997).

\end{thebibliography}
\end{document}